# Multiview Hessian Discriminative Sparse Coding for Image Annotation

W. Liu, D. Tao*, J. Cheng and Y. Tang

Centre for Quantum Computation and Intelligent Systems, University of Technology, Sydney, Jones Street, Ultimo, NSW, 2007, Sydney, Australia

**Abstract**

Sparse coding represents a signal sparsely by using an overcomplete dictionary, and obtains promising performance in practical computer vision applications, especially for signal restoration tasks such as image denoising and image inpainting. In recent years, many discriminative sparse coding algorithms have been developed for classification problems, but they cannot naturally handle visual data represented by multiview features. In addition, existing sparse coding algorithms use graph Laplacian to model the local geometry of the data distribution. It has been identified that Laplacian regularization biases the solution towards a constant function which possibly leads to poor extrapolating power. In this paper, we present multiview Hessian discriminative sparse coding (mHDSC) which seamlessly integrates Hessian regularization with discriminative sparse coding for multiview learning problems. In particular, mHDSC exploits Hessian regularization to steer the solution which varies smoothly along geodesics in the manifold, and treats the label information as an additional view of feature for incorporating the discriminative power for image annotation. We conduct extensive experiments on PASCAL VOC'07 dataset and demonstrate the effectiveness of mHDSC for image annotation.

*Keywords*: Image annotation; Hessian; multiview; sparse coding

# 1. Introduction

Due to the prodigious development of sensors such as cameras and microphones, people can exploit huge amounts of high dimensional data carrying particular kinds of information. Considering the redundancy of these high dimensional data for a particular intelligent task, such as object categorization and human behaviour analytics, it is essential to properly represent the relevant information to reveal the underlying process of these observations.

Sparse coding aims to learn a dictionary and simultaneously find a sparse linear combination of atoms from this dictionary to represent the observations (*e.g.* images and image features). It has received growing attentions because of its flexibility and promising performance for many computer vision applications, such as image denosing [1] and inpainting [4].

In recent years, dozens of sparse coding algorithms have been developed and these algorithms can be grouped into the following five categories: reconstructive sparse coding, supervised sparse coding, discriminative sparse coding, structured sparse coding and graph regularized sparse coding.

> (1) Reconstructive sparse coding: Reconstructive sparse coding methods learn the optimal dictionary and find the corresponding sparse representation by minimizing the data reconstruction error. The representative optimization methods for sparse representation include matching pursuit [16], orthogonal matching pursuit [19] and basis pursuit [3].
>
> (2) Supervised sparse coding: Supervised sparse coding methods exploit the label information to learn an over-completed dictionary and the corresponding sparse representation for classification tasks. Pham *et al.* [20] considered the class label and the linear predictive classification error and proposed a joint framework of dictionary construction and classification. Zhang *et al.* [29] incorporated the labels directly into the sparse coding stage and proposed a discriminative K-SVD (D-KSVD) method to retain the separability. Jiang *et al.* [13] extended D-KSVD by integrating both labels and classification error.

(3) Discriminative sparse coding: In contrast to supervised sparse coding which straightforwardly exploit the class label information, discriminative sparse coding methods incorporate class separability criterion into the objective function. Popular class separability criteria include softmax function [15], Fisher discrimination criterion [24], and hinge loss [18]. Mairal *et al.* [15] used the classical softmax discriminative cost function to leverage the sparse coding. Yang *et al.* [24] introduced Fisher's discriminative criterion to sparse coding to ensure the sparse representations have large between-class scatter but small within-class scatter. Lian *et al.* [18] proposed a max-margin sparse coding method which combined the hinge loss function with sparse coding.

(4) Structured sparse coding: Structured sparse coding methods naturally extend reconstructive sparse coding by exploiting the structure sparsity such as group sparsity [28] and hierarchical sparsity [11]. Yuan and Lin [28] extended Lasso to group Lasso which considered group/block structured dependencies among the sparse coefficients. Jenatton *et al.* [11] employed hierarchical sparsity-inducing norms to learn a hierarchical dictionary which solved tree-structured sparse decomposition problems. Jia *et al.* [12] exploited structured sparsity to learning a latent space of multiview data.

(5) Graph regularized sparse coding: Graph regularized sparse coding methods use graph regularization to exploit the local geometry of the data distribution. Graph Laplacian is a representative graph regularization. Zheng *et al.* [30] used graph Laplacian to exploit the local geometry of the data distribution by adding a Laplacian regularization (LR) to the sparse coding framework. Gao *et al.* [8] proposed hypergraph Laplacian regularized sparse coding to preserve the local consistence in the feature space.

Although the aforementioned sparse coding algorithms have obtained promising performance for various applications such as clustering, classification, and dimensional reduction, they share some of the following two major problems for image annotation:

(1) Since it is expensive to label a large number of images for training a robust model, manifold assumption based semi-supervised learning (SSL) has been introduced to integrate both a small number of labelled images and a large number of unlabelled images to improve the performance of image annotation. LR is one of the most representative works in which the geometry of the underlying manifold is determined by the graph Laplacian. Although LR achieved top level performance for image annotation, it suffers from lacking of extrapolating power. It has been identified that LR biases the solution towards a constant function due to its constant null space, which possibly leads to poor extrapolation capability [14].

(2) The aforementioned sparse coding methods are only applicable to images that are represented by single view features. However, in image annotation, images are often described by multiview features. Different views (or equivalently visual features), such as colour histogram, edge sketch and local binary patterns (LBP), characterize different properties of an image [7,17,21]. Each view of a feature describes a specific property of the image, and the weaknesses of a particular view can be reduced by the strengths of others. Although we can concatenate different features into a long vector, this concatenation strategy cannot efficiently explore the complementary of different features because it improperly treats different features carrying different physical characteristics. Therefore, compared to single view feature, multiview features provide more characteristics of images and can significantly leverage the performance especially when features for different views are complementary to one another.

To address these problems, we present multiview Hessian discriminative sparse coding (mHDSC) in this paper. Particularly, mHDSC can well leverage multiview sparse coding by seamless integrating Hessian regularization with discrimination. According to proposition 1 in [14], the geodesic function in null space of Laplacian is no other than a const, which implicates that LR biases the solution towards a constant function and then leads to poor extrapolation capability. In contrast to Laplacian, Hessian has richer null space and drives the solution varying smoothly along the manifold. Hessian regularization (HR) is more preferable for exploiting the local geometry than LR. Kim et al. [14] has demonstrated the excellent performance of HR in regression problems. The proposed mHDSC has the following advantages: (1) mHDSC incorporates multiview features into sparse coding, which effectively explores the complementation of different features from different views; (2) mHDSC treats the label information as an additional view of feature, which well boosts the discrimination without adding more computing complexity; and (3) mHDSC exploits Hessian regularization to preserve local similarity, which steers the solution varying smoothly along geodesics in the manifold.

We carefully implement mHDSC for image annotation and conduct experiments on the PASCAL VOC'07 dataset [6]. To evaluate the performance of mHDSC, we also compare mHDSC with several baseline algorithms including discriminative sparse coding (DSC), Laplacian discriminative sparse coding (LDSC), Hessian discriminative sparse coding (HDSC), multiview sparse coding (mSC), multiview discriminative sparse coding (mDSC) and multiview Laplacian discriminative sparse coding (mLDSC). The experimental results demonstrate the effectiveness of mHDSC by comparison with the baseline algorithms.

The rest of this paper is arranged as follows. Section 2 presents the proposed mHDSC framework. Section 3 details the implementation of mHDSC. Section 4 discusses some related work. And Section 5 demonstrates experimental results followed by the conclusion in section 6.

## 2. multiview Hessian discriminative sparse coding

In multiview sparse coding (mSC), we are given a multiview dataset of $N$ observations from $V$ views including $l$ labelled data i.e. $S_L = \{x_i^{(1)}, x_i^{(2)}, \ldots, x_i^{(V)}, y_i\}_{i=1}^{l}$ and $u$ unlabelled data i.e. $S_U = \{x_i^{(1)}, x_i^{(2)}, \ldots, x_i^{(V)}\}_{i=l+1}^{N}$, where $y_i \in R^{P_c}$ is the class labels of the $i^{th}$ example ($P_c$ is the number of class). In the following section of this paper, we use $X_L^{(v)} \in R^{P_v \times l}$ to denote the $v^{th}$ view feature vectors of labelled data ($P_v$ is the dimension of the $v^{th}$ view feature), $Y \in R^{P_c \times l}$ to denote the label vectors, and $X_U^{(v)} \in R^{P_v \times (N-l)}$ to denote the $v^{th}$ view feature vectors of unlabelled data.

By incorporating an additional regularization term to control the sparsity and exploit the local geometry, mSC aims to find an integrated sparse representation (code) $W \in R^{N_d \times N}$ of the multiview data and a multiview dictionary $D = \{D^{(1)}, D^{(2)}, \ldots, D^{(V)}\}$, where $D^{(v)} \in R^{P_v \times N_d}$ contains $N_d$ dictionary atoms for the view $v$. Thus, mSC is written as follows

$$\min_{D,W} \frac{1}{2N} \sum_{v=1}^{V} \|X^{(v)} - D^{(v)}W\|_F^2 + \varphi(W), \tag{1}$$

$$s.t. \|D_i^{(v)}\| \leq 1, 1 \leq i \leq N_d, X^{(v)} = \{X_L^{(v)}, X_U^{(v)}\},$$

where $\varphi(W) = \gamma_1 \varphi_1(W) + \gamma_2 \varphi_2(W) + \gamma_3 \varphi_3(W)$, $\varphi_1(W) = \|W\|_{1,\infty}$ is a regularizer that controls the sparsity over $W$, $\varphi_2(W) = \sum_{v=1}^{V} \|(D^{(v)})^T\|_{1,\infty}$ is a regularizer that controls the structure of dictionary, $\varphi_3(W)$ is a regularizer to preserve the local similarity, and $\gamma_1$, $\gamma_2$ and

$\gamma_3$ are parameters that balance the loss function and regularizations $\varphi_1(W)$, $\varphi_2(W)$ and $\varphi_3(W)$, respectively.

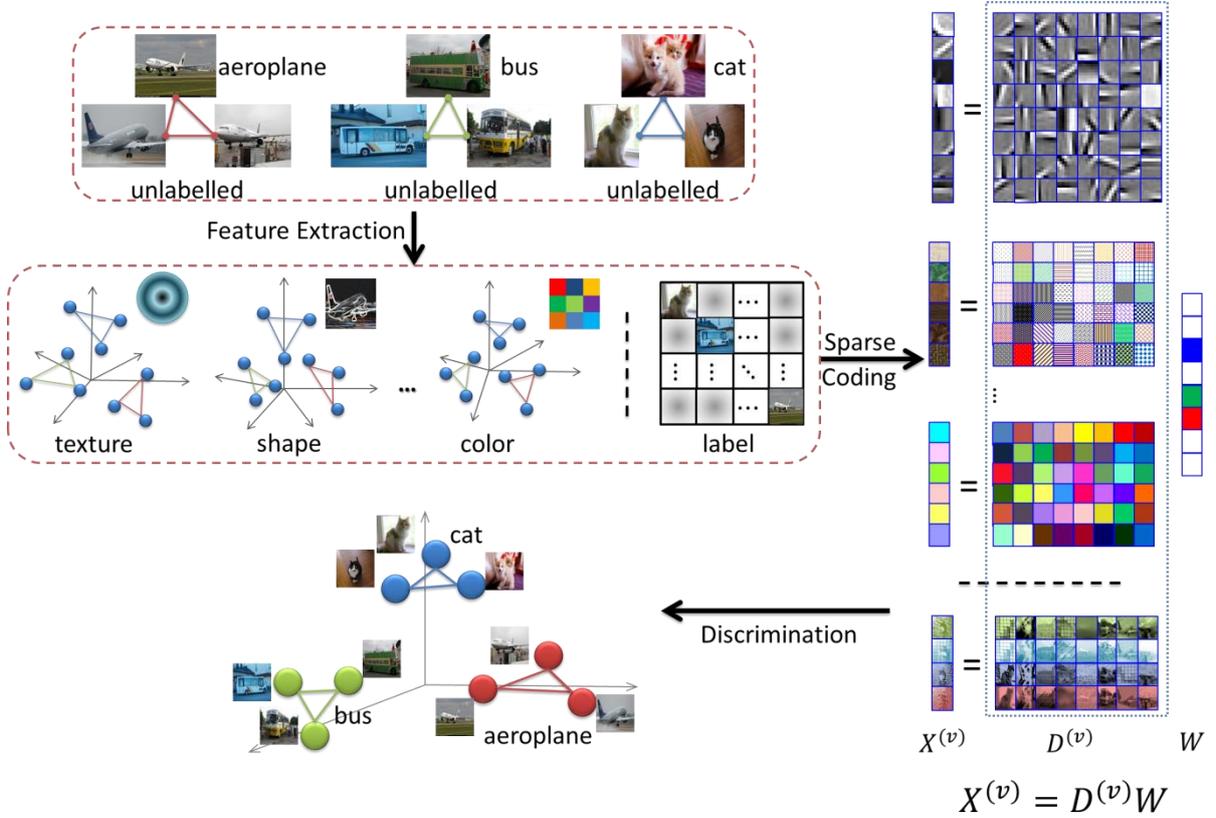

Fig. 1. mHDSC framework.

Although there are different choices for $\varphi_2(W)$ to exploit the local geometry, Laplacian regularization (LR) [30, 8] is promising to preserve the local similarity. It is crucial to accurately explore the local geometry in semi-supervised image annotation, because images share similar semantic concepts should be close in the representation w.r.t. the new bases (i.e. the multiview dictionary). Thus, the corresponding sparse codes of the images which share common labels are close to each other. However, LR biases the solution towards a constant function [14], so it is not the best choice for encoding the local geometry for semi-supervised image annotation.

In this paper, we propose multiview Hessian discriminant sparse coding (mHDSC) for image annotation. Fig. 1 describes the framework of mHDSC. Particularly, mHDSC employs Hessian regularization (HR) to encode the local geometry. And it is applied to multiview features. In addition, mHDSC treats the label information as an additional view of feature to boost the discrimination of the dictionary. Thus mHDSC can be expressed as follows according to (1)

$$\min_{D,W,\alpha} \frac{1}{2l}\sum_{v=1}^{V+1}\left\|X_L^{(v)} - D^{(v)}W_L\right\|_F^2 + \frac{1}{2(N-l)}\sum_{v=1}^{V}\left\|X_U^{(v)} - D^{(v)}W_U\right\|_F^2 + \gamma_1\|W\|_{1,\infty} +$$

$$\gamma_2 \sum_{v=1}^{V+1}\left\|\left(D^{(v)}\right)^T\right\|_{1,\infty} + \gamma_3 tr(W(\sum_{v=1}^{V+1}(\alpha_v H_v))W^T) \quad (2)$$

$$s.t. \left\|D_i^{(v)}\right\|^2 \leq 1, 1 \leq i \leq N_d, \sum_{v=1}^{V+1}\alpha_v = 1, \alpha_v \geq 0, W = \{W_L, W_U\},$$

where $X_L^{(V+1)} = Y$ is the label information, $W_L$ is the sparse representation of labelled data, $W_U$ is the sparse representation of unlabelled data, and $H_v$ is the Hessian regularization computed from the $v^{th}$ view feature. It has been shown by Kim [14] that Hessian regularization improves the performance of LR and steers the solution varying smoothly along the coordinate system.

Although there are several ways to define the loss function in (2), we use the traditional least squares loss of sparse coding due to its efficiency and simplicity. This loss has been widely adopted in practice, such as [31].

The proposed mHDSC treats $Y$ as the additional view feature, so it can infer the label information from the sparse code without using classifiers. In particular, given a test image represented by multiview features $x^{(v)}, 1 \leq v \leq V$, mHDSC can estimate the label $y$, i.e. the $(v+1)^{th}$ view feature by conducting the following two steps. It first obtains $w$ by solving a convex problem

$$\min_w \frac{1}{2}\sum_{v=1}^{V}\left\|x^{(v)} - D^{(v)}w\right\|_F^2 + \gamma_1\|w\|_1. \quad (3)$$

Then the label $y$, i.e. the $(v+1)^{th}$ view feature, is given by

$$y = x^{(V+1)} = D^{(V+1)}w.$$

In contrast to existing works [32,33,34,35] that exploit either sparse learning-based feature selection or multi-feature fusion for image annotation and video retrieval, the proposed mHDSC learns a dictionary while finding the sparse linear combination of atoms from this dictionary to represent the multiple observations. Especially, the proposed mHDSC has the following advantages: (1) it naturally encodes the discrimination by treating labels as an additional view feature; (2) it exploits the complementary property of different visual features by incorporating multiview features into the sparse coding framework; (3) it precisely captures the second order information of the local geometry by exploiting Hessian regularization to preserve local geometry; and (4) it can infer the label information from the sparse code without using classifiers.

The objective function in (2) is convex w.r.t. $D$, $W$ or $\alpha$ seperately, but not jointly convex w.r.t. $D$, $W$ and $\alpha$. In this paper, we iteratively solve the problem by using alternating optimization [2] which optimizes over one variable with others fixed. Generally, the solution of (2) can be divided into three parts: sparse coding, dictionary updating and graph coefficients updating. In section 3, we detail the optimization algorithm of (2).

## 3. Algorithms

The optimization of mHDSC contains three steps: (1) learning sparse codes W given fixed dictionary D and graph coefficients α; (2) updating dictionary D given fixed sparse codes W and graph coefficients α; (3) learning optimal graph coefficients α given fixed sparse codes W and dictionary D. In the following, we first provide a brief description of the alternating optimization of mHDSC, and then present the optimization of each subproblem in detail. For convenience, Table 1 lists the important notations used in this paper.

Table 1. List of important notations

| Notation | Description | Notation | Description |
| --- | --- | --- | --- |
| $l$ | number of labelled data | $N$ | number of data |
| $X_L^{(v)}$ | the $v^{th}$ view of labelled examples | $X_U^{(v)}$ | the $v^{th}$ view of unlabelled examples |
| $D^{(v)}$ | dictionary of the $v^{th}$ view | $V$ | number of views |
| $N_d$ | the number of dictionary atoms | $W$ | $W = \{W_L, W_U\}$ |
| $W_L$ | sparse codes of labelled data | $W_U$ | sparse codes of unlabelled data |
| $H_v$ | Hessian regularization w.r.t $v^{th}$ view feature. | $\boldsymbol{H}$ | multiview Hessian, $H = \sum_{v=1}^{V+1}(\alpha_v H_v)$ |
| $\alpha$ | graph coefficients | $\gamma_1$ | parameter of $\|W\|_{1,\infty}$ |
| $\gamma_2$ | parameter of $\sum_{v=1}^{V+1}\left\|(D^{(v)})^T\right\|_{1,\infty}$ | $\gamma_3$ | parameter of $tr(W(\sum_{v=1}^{V+1}(\alpha_v H_v))W^T)$ |

Given fixed $D$ and $\alpha$, the problem (2) can be simplified to:

$$\min_W \frac{1}{2l}\sum_{v=1}^{V+1}\left\|X_L^{(v)} - D^{(v)}W_L\right\|_F^2 + \frac{1}{2(N-l)}\sum_{v=1}^{V}\left\|X_U^{(v)} - D^{(v)}W_U\right\|_F^2 + \gamma_1\|W\|_{1,\infty} + \gamma_3 tr(W\boldsymbol{H}W^T), \quad (4)$$

where $\boldsymbol{H} = \sum_{v=1}^{V+1}(\alpha_v H_v), \sum_{v=1}^{V+1}\alpha_v = 1, \alpha_v \geq 0$.

Given fixed $W$ and $\alpha$, the problem (2) can be simplified to:

$$\min_D \frac{1}{2l}\sum_{v=1}^{V+1}\left\|X_L^{(v)} - D^{(v)}W_L\right\|_F^2 + \frac{1}{2(N-l)}\sum_{v=1}^{V}\left\|X_U^{(v)} - D^{(v)}W_U\right\|_F^2 + \gamma_2 \sum_{v=1}^{V+1}\left\|(D^{(v)})^T\right\|_{1,\infty},$$

$$(5)$$

$$s.t. \left\|D_i^{(v)}\right\|^2 \leq 1, 1 \leq i \leq N_d.$$

And given fixed $D$ and $W$, the problem (2) can be simplified to:

$$\min_\alpha \gamma_3 tr(W(\sum_{v=1}^{V+1}(\alpha_v H_v))W^T), \tag{6}$$

$$s.t. \sum_{v=1}^{V+1} \alpha_v = 1, \alpha_v \geq 0.$$

Algorithm 1 summarizes the overall procedure of the alternating optimization.

**Algorithm 1.** Alternating optimization method for (2).

---

**Input**: $X_L, X_U, Y, \gamma_1, \gamma_2, \gamma_3$

**Output**: $D, W, \alpha$

1: Initialize $D, W, \alpha$, *e.g.*, with random entries.

2: **repeat**

3:  Update $W$ as

$$W \leftarrow arg\min_W \frac{1}{2l}\sum_{v=1}^{V+1}\left\|X_L^{(v)} - D^{(v)}W_L\right\|_F^2 + \frac{1}{2(N-l)}\sum_{v=1}^{V}\left\|X_U^{(v)} - D^{(v)}W_U\right\|_F^2 + \gamma_1\|W\|_{1,\infty} + \gamma_3 tr(WHW^T).$$

4:  Update $D$ as

$$D \leftarrow arg\min_D \frac{1}{2l}\sum_{v=1}^{V+1}\left\|X_L^{(v)} - D^{(v)}W_L\right\|_F^2 + \frac{1}{2(N-l)}\sum_{v=1}^{V}\left\|X_U^{(v)} - D^{(v)}W_U\right\|_F^2 + \gamma_2 \sum_{v=1}^{V+1}\left\|\left(D^{(v)}\right)^T\right\|_{1,\infty}.$$

5:  Update $\alpha$ as

$$\alpha \leftarrow arg\min_\alpha \gamma_3 tr(W(\sum_{v=1}^{V+1}(\alpha_v H_v))W^T).$$

6: **until** convergence

---

3.1 Learning sparse codes $W$

The subproblem (4) can be written as the following general form:

$$\min_W f(W) + g(W) \tag{7}$$

where $f(W) = \frac{1}{2l}\sum_{v=1}^{V+1}\left\|X_L^{(v)} - D^{(v)}W_L\right\|_F^2 + \frac{1}{2(N-l)}\sum_{v=1}^{V}\left\|X_U^{(v)} - D^{(v)}W_U\right\|_F^2 + \gamma_3 tr(WHW^T)$,

$g(W) = \gamma_1\|W\|_{1,\infty}$. Both $f(W)$ and $g(W)$ are convex functions. Furthermore, $f(W)$ is differentiable and $\nabla f(W)$ is Lipschitz continuous. Hence we can adopt an efficient convex optimization method for subproblem (4). Algorithm 2 shows a variant of Nesterov's first order method which can solve (4).

**Algorithm 2.** A convex optimization method for (4).

---

**Input**: $X_L, X_U, Y, D = \{D_1, D_2\}, D_2 \in R^{P_c \times N_d}$ $\boldsymbol{H}, \gamma_1, \gamma_3$.

**Output**: $W$

1: Choose $W^{(0)}, \widetilde{W}^{(0)}$ and let $\tau^{(0)} = 1$ and $L_1 = \frac{1}{l}\sigma_{max}(D^T D), L_2 = \frac{1}{N-l}\sigma_{max}(D_1^T D_1), L_3 = \gamma_3 \sigma_{max}(\boldsymbol{H})$.

2: **for** $k=0,1,2,\ldots$, until convergence **do**

3: $\quad Z^{(k)} = \{Z_L^{(k)}, Z_U^{(k)}\} \leftarrow \tau^{(k)} W^{(k)} + (1-\tau^{(k)})\widetilde{W}^{(k)}$

4: $\quad$ Update

$$W_L^{(k+1)} \leftarrow \operatorname{argmin}_{W_L} \left\| W_L - U_L^{(k)} \right\|_F^2 + \frac{\gamma_1}{\tau^{(k)}(L_1+L_3)} \|W_L\|_{1,\infty},$$

$$W_U^{(k+1)} \leftarrow \operatorname{argmin}_{W_U} \left\| W_U - U_U^{(k)} \right\|_F^2 + \frac{\gamma_1}{\tau^{(k)}(L_2+L_3)} \|W_U\|_{1,\infty}, \quad (8)$$

where $U_L^{(k)} = W_L^{(k)} - \frac{1}{\tau^{(k)}(L_1+L_3)}\left(\frac{1}{l}\left(D^T D Z_L^{(k)} - D^T X_L\right) + \gamma_3 (W\boldsymbol{H})_L\right)$ and

$U_U^{(k)} = W_U^{(k)} - \frac{1}{\tau^{(k)}(L_2+L_3)}\left(\frac{1}{N-l}\left(D_1^T D_1 Z_U^{(k)} - D_1^T X_U\right) + \gamma_3 (W\boldsymbol{H})_U\right).$

5: $\quad \widetilde{W}^{(k+1)} = \tau^{(k)} W^{(k+1)} + (1-\tau^{(k)})\widetilde{W}^{(k)}$

6: $\quad$ Find $\tau^{(k+1)} > 0$ such that

$$\left(\tau^{(k+1)}\right)^{-2} - \left(\tau^{(k+1)}\right)^{-1} = \left(\tau^{(k)}\right)^{-2}$$

7: **end for**

8: return $\widetilde{W}^{(k)}$

---

Subproblem (8) can be separated w.r.t. each row of $W$ and efficiently solved using $l_1$ projection [5].

### 3.2 Updating dictionary $D$

Similarly, the subproblem (5) can be separated to $(V+1)$ parts w.r.t. each view. And each part also can be written as the general form of (7) with $f(W) = \frac{1}{2l}\left\|X_L^{(v)} - D^{(v)}W_L\right\|_F^2 + \frac{1}{2(N-l)}\left\|X_U^{(v)} - D^{(v)}W_U\right\|_F^2$ and $g(W) = \gamma_2 \left\|\left(D^{(v)}\right)^T\right\|_{1,\infty}$. Therefore we can also adopt the framework of Algorithm 2 to solve the subproblem (5). We brief the optimization of (5) in Algorithm 3.

**Algorithm 3.** A convex optimization method for (5).

---

**Input**: $X_L, X_U, Y, W, \gamma_2$

**Output**: $D = \{D^{(1)}, D^{(2)}, \dots, D^{(V+1)}\}$

For each $v$, denote $\left(D^{(v)}\right)^T$ as $B$

1:    Choose $B^{(0)}, \tilde{B}^{(0)}$ and let $\tau^{(0)} = 1$ and $L_1 = \frac{1}{l}\sigma_{max}(W_L W_L^T), L_2 = \frac{1}{N-l}\sigma_{max}(W_U W_U^T)$.

2:    **for** $k=0,1,2,\dots$, until convergence **do**

3:       $Z^{(k)} = \left\{Z_L^{(k)}, Z_U^{(k)}\right\} \leftarrow \tau^{(k)}B^{(k)} + \left(1 - \tau^{(k)}\right)\tilde{B}^{(k)}$

4:       Update

        case: $v = 1, \dots, V$

$$B^{(k+1)} \leftarrow \operatorname{argmin}_B \left\|B - U^{(k)}\right\|_F^2 + \frac{\gamma_1}{\tau^{(k)}(L_1+L_2)}\|B\|_{1,\infty},$$

where $U^{(k)} = B^{(k)} - \frac{1}{\tau^{(k)}(L_1+L_2)}\left(\frac{1}{l}\left(W_L W_L^T Z_L^{(k)} - W_L X_L^T\right) + \frac{1}{N-l}\left(W_U W_U^T Z_U^{(k)} - W_U X_U^T\right)\right)$.

        case: $v = V + 1$

$$B^{(k+1)} \leftarrow \operatorname{argmin}_B \left\|B - U^{(k)}\right\|_F^2 + \frac{\gamma_1}{\tau^{(k)}L_1}\|B\|_{1,\infty},$$

where $U^{(k)} = B^{(k)} - \frac{1}{\tau^{(k)}L_1}\left(\frac{1}{l}\left(W_L W_L^T Z_L^{(k)} - W_L X_L^T\right)\right)$.

---

5: $\quad \tilde{B}^{(k+1)} = \tau^{(k)} B^{(k+1)} + \left(1 - \tau^{(k)}\right) \tilde{B}^{(k)}$

6: $\quad$ Find $\tau^{(k+1)} > 0$ such that

$$\left(\tau^{(k+1)}\right)^{-2} - \left(\tau^{(k+1)}\right)^{-1} = \left(\tau^{(k)}\right)^{-2}$$

7: **end for**

8: $\quad \left(D^{(v)}\right)^T \leftarrow \tilde{B}^{(k)}$

9: return $D$

### 3.3 Learning graph coefficients $\alpha$

The subproblem (6) can be equally rewritten as

$$\min_\alpha \sum_{v=1}^{V+1} \alpha_v \, tr(WH_v W^T), \tag{9}$$

$$s.t. \sum_{v=1}^{V+1} \alpha_v = 1, \alpha_v \geq 0.$$

The solution w.r.t. $\alpha$ is $\alpha_i = 1$ when $tr(WH_i W^T)$ is the minimum one over different views, and $\alpha_k = 1$ otherwise. This means that only one view is selected and this method cannot explore the complementary property of multiple views.

In this paper, we employ a trick [22, 23] to avoid this phenomenon, *i.e.* we replace $\alpha_i$ with $\alpha_i^r, r > 1$. Under this setting, each view has a particular contribution to the final sparse coding. And therefore, the new objective function of (9) is expressed as:

$$\min_\alpha \sum_{v=1}^{V+1} \alpha_v^r \, tr(WH_v W^T), \tag{10}$$

$$s.t. \sum_{v=1}^{V+1} \alpha_v = 1, \alpha_v \geq 0.$$

To solve (10), let $\lambda$ be a Lagrange multiplier and consider the constraint $\sum_{v=1}^{V+1} \alpha_v = 1$, and then we get the Lagrange function

$$L(\alpha, \lambda) = \sum_{v=1}^{V+1} \alpha_v^r tr(WH_v W^T) - \lambda\left(\sum_{v=1}^{V+1} \alpha_v - 1\right). \tag{11}$$

By setting the derivative of $L(\alpha, \lambda)$ w.r.t. $\alpha_v$ and $\lambda$ to zero, we have

$$\begin{cases} \frac{\partial L(\alpha,\lambda)}{\partial \alpha_v} = r\alpha_v^{r-1} tr(WH_vW^T) - \lambda = 0, v = 1, \dots, V+1 \\ \frac{\partial L(\alpha,\lambda)}{\partial \lambda} = \sum_{v=1}^{V+1} \alpha_v - 1 = 0. \end{cases} \quad (12)$$

Therefore, a closed form solution $\alpha_v$ can be obtained

$$\alpha_v = \frac{\left(1/tr(WH_vW^T)\right)^{1/(r-1)}}{\sum_{v=1}^{V+1}\left(1/tr(WH_vW^T)\right)^{1/(r-1)}}. \quad (13)$$

The Hessian matrix $H_v$ is semi-definite positive, and thus we always have $\alpha_v \geq 0$. When $W$ is fixed, (13) gives the global optimal $\alpha$.

3.4 Complexity Analysis

Suppose we are given n samples, v view features. Denote the number of dictionary D atoms as d, the sum dimension of all view features as p, and the number of iteration as k for subproblem (4) and (5), we optimize W and D with the time complexity $kO(d^2(n+p) + dpn + dn^2)$ and $kO(d^2(n+p) + dpn)$, respectively. And the time cost for subproblem (6) is $O((v+d) \times n^2)$. Denote the number of alternating iterations as η, and the number of candidate parameters that need the m-fold cross-validation as r. Therefore, the total cost of the proposed method is $O(mηr(2kd^2(n+p) + 2kdpn + (kd+v+d)n^2))$. Since the view number v and dictionary atom number d is generally much smaller than the product of d and iteration number k, the time cost is approximately $O(mηr(2kd^2(n+p) + 2kdpn + kdn^2))$. When d is much smaller than n and p, the time cost is around $O(mηr(2kdpn + kdn^2))$. When the image set becomes larger (i.e. $n \gg p$), the time cost can be approximate as $O(mηr(kdn^2))$. Since matrix product cost most of the computational time in the proposed method, parallelization (e.g. MapReduce and GPU computing) can be employed to efficiently reduce the time cost.

## 4. Related work

Suppose we are given a set of $N$ observations with the corresponding labels, *i.e.* $S = \{X, Y\}$, where $X \in R^{P \times N}$ contains $N$ feature vectors each is of dimensionality $P$ and the label matrix $Y \in R^{P_c \times N}$ contains the corresponding label vectors.

Sparse coding aims to learn a sparse code $W \in R^{N_d \times N}$ and a dictionary $D \in R^{P \times N_d}$. Mathematically, it can be written as

$$\min_{D,W} \frac{1}{2N} \|X - DW\|_F^2 + \varphi(W),$$

where $\varphi(W)$ is a regularizer over $W$ to control the volume of the search space of $W$. According to different motivations and purposes, we can use different forms of $\varphi(W)$. Examples are given below.

**Reconstructive sparse coding** [16, 19, 3] uses $\varphi(W) = \gamma \sum_{i=1}^{N} \|W_i\|_1$ to make $W$ sparse.

**Supervise sparse coding** [20] encodes the label information for constructing $\varphi(W)$

$$\varphi(W) = \gamma_1 \sum_{i=1}^{N} \|W_i\|_1 + \gamma_2 \|Y - AW\|_F^2 + \gamma_3 \|A\|_F^2,$$

where $A$ is the parameter of linear predictive classifier, $Y$ is the label information and $\gamma_1, \gamma_2, \gamma_3$ are the parameters to balance the regularizer terms.

**Discriminative sparse coding** does not straightforwardly exploit the label information but develops a particular discriminative item to convey the label information. Different discriminative sparse coding algorithms consider different discriminative terms.

Mairal *et al.* [15] considered the softmax discriminative cost function

$$\varphi(W) = \gamma_1 \sum_{i=1}^{N} \|W_i\|_1 + \mathcal{C}(W, D),$$

where $\mathcal{C}(W, D)$ is the softmax cost function [15].

Yang *et al.* [24] considered Fisher's discriminative information

$$\varphi(W) = \gamma_1 \sum_{i=1}^{N} \|W_i\|_1 + \gamma_2 (tr(S_W(W) - S_B(W)) + \gamma_3 \|W\|_F^2),$$

where $S_W(W)$ is the within-class scatter of $W$ and $S_B(W)$ is the between-class scatter of $W$, and $\gamma_1, \gamma_2, \gamma_3$ are the parameters.

Lian *et al.* [18] considered the max-margin information

$$\varphi(W) = \gamma_1 \sum_{i=1}^{N}\|W_i\|_1 + \gamma_2 \|A\|_2^2 + \gamma_3 \sum_{i=1}^{N} \max(0, 1 - Y_i < A, \emptyset >),$$

where $A$ is the hyper-plane classifier, $\emptyset$ is a data descriptor based on $W$, and $\gamma_1, \gamma_2, \gamma_3$ are the parameters.

**Structured sparse coding** exploits the structure sparsity over the codes. Different structured sparse coding algorithms consider different structure sparsity terms.

Yuan et al. [28] considered group sparsity

$$\varphi(W) = \gamma \sum_{i=1}^{N} \sum_{j=1}^{J} \left\|W_i^{(j)}\right\|_{K_j},$$

where $K_j$ is a symmetric positive definite matrix for group selection, and $\|\eta\|_{K_j} = (\eta^T K_j \eta)^{1/2}$ is the induced norm that makes intermediate regularization between $l_1$ and $l_2$.

Jenatton *et al.* [11] considered tree-structured sparse regularization

$$\varphi(W) = \gamma \sum_{i=1}^{N} \Omega(W_i),$$

where $\Omega(\eta)$ is a hierarchical sparse-inducing norm that leads to a tree-structure of sparse codes [11].

Jia *et al.* [12] used sparse coding techniques to factorize multiple representations.

$$\min_{D,W} \frac{1}{2N} \sum_{v=1}^{V+1} \left\|X^{(v)} - D^{(v)}W\right\|_F^2 + \gamma_1 \|W\|_{1,\infty} + \gamma_2 \sum_{v=1}^{V} \left\|\left(D^{(v)}\right)^T\right\|_{1,\infty}.$$

**Graph regularized sparse coding** exploits the local geometry of the data distribution. Different graph regularized sparse coding algorithms consider different graph regularizations.

Zheng *et al.* [30] considered Laplacian regularization

$$\varphi(W) = \gamma_1 \sum_{i=1}^{N}\|W_i\|_1 + \gamma_2 tr(WLW^T),$$

where $L$ is the Laplacian matrix, and $\gamma_1, \gamma_2$ are the parameters.

Gao *et al.* [8] considered hypergraph Laplacian regularization

$$\varphi(W) = \gamma_1 \sum_{i=1}^{N}\|W_i\|_1 + \gamma_2 tr(WL^hW^T),$$

where $L^h$ is the Hyperlaplacian matrix, and $\gamma_1, \gamma_2$ are the parameters.

In contrast to the aforementioned works, the proposed mHDSC can (1) properly explore the complementation of multiview features for sparse coding; (2) well boost the discrimination by simply incorporating the label information into sparse coding framework; and (3) effectively encode the local geometry by using the Hessian regularization.

## 5. Experiments

To evaluate the effectiveness of the proposed mHDSC, we apply linear SVM classifier and least squares (LS) to the integrated sparse codes obtained by mHDSC for image annotation [10], respectively. We also combine the sparse codes and the learned dictionary to infer the label information for image annotation. We conduct the experiments on the PASCAL VOC'07 dataset [6] which contains 9,963 images of 20 visual object classes. Fig. 2 shows example images of 6 classes.

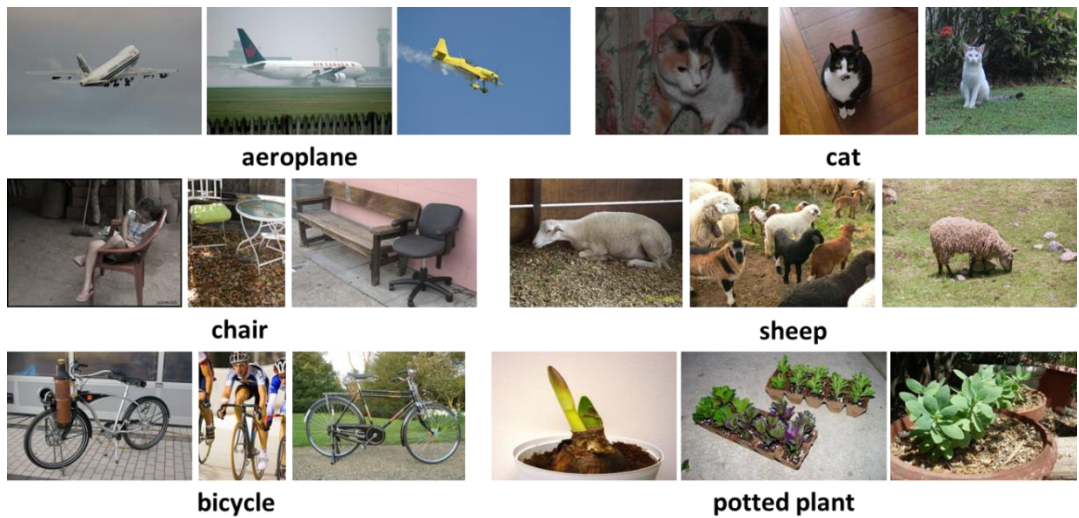

Fig. 2. Example images of the PASCAL VOC'07 database.

We use 15 visual features provided by Guillaumin et al. [9] including a GIST feature, 6 colour features (2 RGB features, 2 Lab features and 2 HSV features), 2 Hue features, 2 SIFT features, 2 Harris+Hue features, and 2 Harris+SIFT features. In particular, global colour histograms are obtained from the RGB, HSV, and LAB colour spaces. Local SIFT features and local Hue histograms are computed on a dense grid and on the Harris interest-point detected regions, respectively. Then the quantized local descriptors are represented by a visual word histogram (e.g. "Rgb", "DenseHue", and "HarrisSift"). And a new histogram representation which encodes spatial information on each histogram is constructed by computing over a $3 \times 1$ horizontal decomposition of the image (e.g. "RgbV3H1", "SDenseHueV3H1", and "HarrisSiftV3H1"). We normalized each view feature in the experiment and the empirical results suggested the uniform weights for the normalized views. According to [6], we divide the dataset into two subsets: a training set which contains 5,011 images and a test set which contains 4,952 images. We further divide the training set into two parts, one contains 4,500 images for training and one contains 511 images for parameter tuning. In the semi-supervised learning experiments, we also assign 10%, 20%, 30%, 50% as labelled data and the rest as unlabelled data. All the parameters are tuned in the case of 50% labelled data and 50% unlabelled data.

The learned sparse codes are input to classifiers (SVM or LS) to conduct annotation. We compare the proposed mHDSC with related sparse coding algorithms which are discriminative sparse coding (DSC), Laplacian discriminative sparse coding (LDSC), Hessian discriminative sparse coding (HDSC), multiview sparse coding (mSC), multiview discriminative sparse coding (mDSC) and multiview Laplacian discriminative sparse coding (mLDSC). The working mechanisms of these algorithms are detained in Section 4. We also compare mHDSC with the feature concatenation method (by concatenating 15 different features into a long feature vector). For all methods, parameters $\gamma_1$, $\gamma_2$ and $\gamma_3$ are tuned from the candidate set $\{1 \times 10^e | e = -10, -4, ..., 10\}$, the number of the dictionary atoms for single view methods and concatenation methods is set to 200, the number of the neighbours in computing Hessian and graph Laplacian is fixed to 100 and r is fixed to 5 empirically.

In our experiments, we measure the performance by using the average precision (AP) and mean average precision (mAP). Particularly, AP and mAP are computed by using the PASCAL VOC method [6]

$$AP = \frac{1}{11}\sum_t [\max_{k \geq t} p(k)], t \in \{0, 0.1, 0.2, \ldots, 1.0\},$$

and

$$mAP = \frac{\sum_{i=1}^{\#} AP_i}{\#\{visual\ object\ classes\}}$$

where $p(k)$ is s the measured precision at recall $k$.

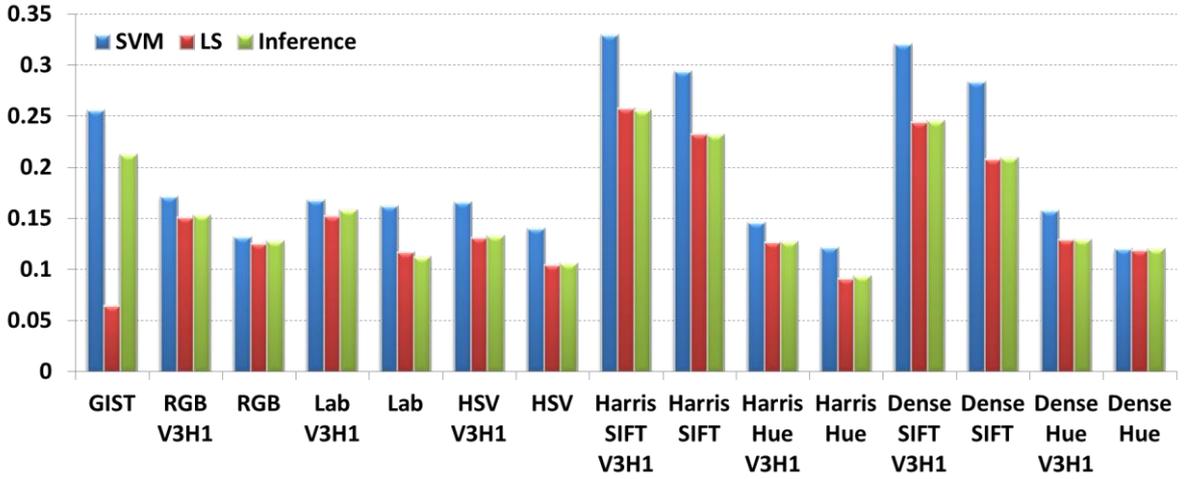

Fig. 3. The mAP of different views.

Multiview Hessian Discriminative Sparse Coding for Image Annotation

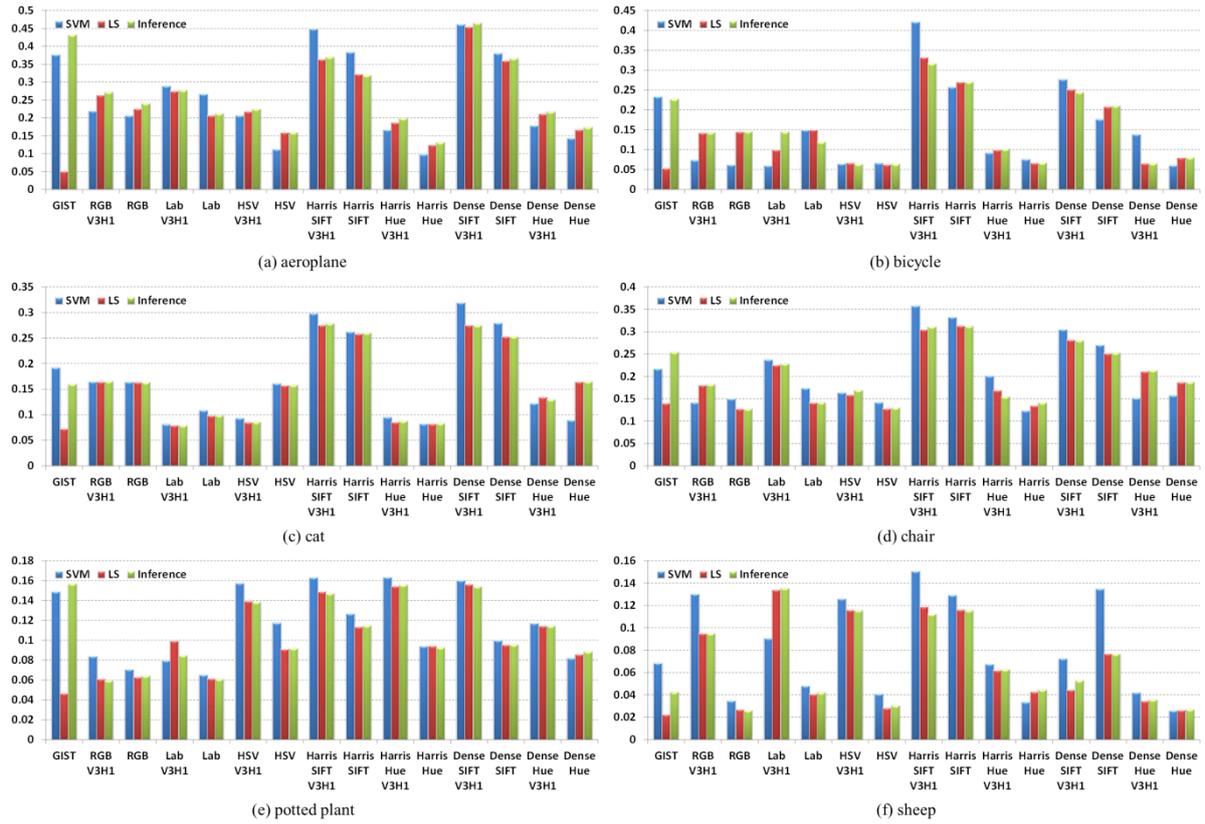

Fig. 4. The AP of different views on selected classes.

## 5.1 Performance of different views

We apply HDSC to each view of feature to evaluate the performance of different views. Fig. 3 shows the mAP of different views. Fig. 4 shows the AP of different views over selected visual object class. From Fig. 3 and Fig. 4, we can see that different visual representations achieve different performances. In the following section, we only consider the single view of the best mAP performance for comparison with multiview ones.

## 5.2 Multiview performance

We compare the performance of mHDSC with different multiview methods and the corresponding best single view ones. Fig. 5 is the mAP of different methods. The subfigures correspond to the performance on different numbers (450, 900, 1350, 2250, 4500) of labelled data. In Fig. 5, BDSC, BLDSC, and BHDSC denote the best single view performance of DSC, LDSC and HDSC methods respectively, CDSC, CLDSC and CHDSC denote the corresponding performance of concatenation methods, and mSC denote the multiview sparse coding that doesn't use label information. From Fig. 5, we can see that multiview discriminative sparse coding methods are better than single view ones and mHDSC outperforms other multiview methods. We can also see that the performance of the inference methods is comparable to that of the LS methods.

Fig. 6 is the AP of different methods over selected visual object class. Each subfigure corresponds to one evaluation method (SVM, LS, Inference) over one visual object class of the selected 6 classes. The x-coordinate is the number of labelled data. From Fig. 6, we can see that multiview methods significantly boost the performance especially when the number of labelled data is small. And mHDSC outperforms other multiview methods.

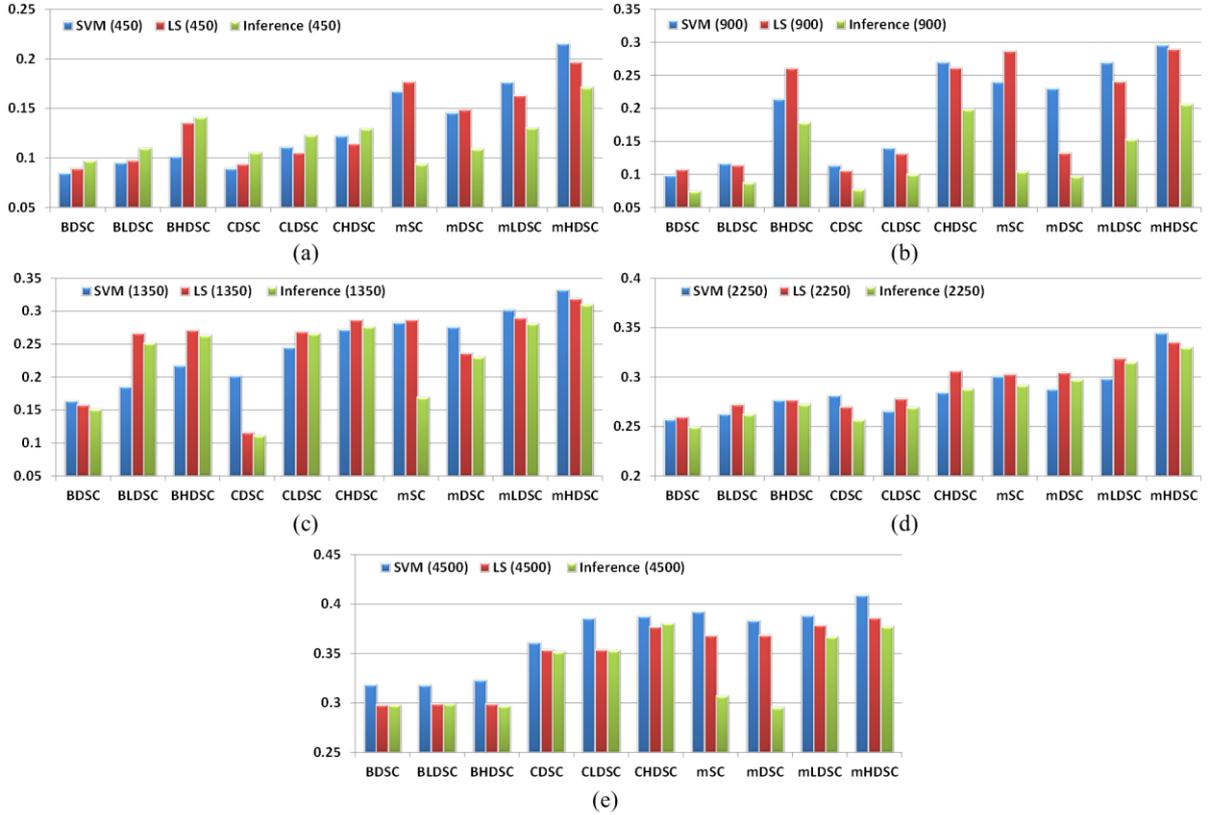

Fig. 5. The mAP of different methods.

## 6. Conclusion

Sparse coding has attracted intensive attentions and achieved promising performance in image annotation. The most prominent sparse coding methods are based on LR. However, LR based sparse coding methods suffer from the poor generalization because LR biases the solution towards a constant function. On the other hand, the current sparse coding methods only deal with single views although images are often represented by multiple visual features. To solve the above two problems, we present multiview Hessian discriminative sparse coding (mHDSC) for image annotation, in which images are represented by multiple visual features. The proposed mHDSC can well explore the local geometry of the data distribution with the help of Hessian regularization and properly utilize the complementary information of multiview features to boost the learning performance. We apply mHDSC to linear SVM and LS regression for image annotation. Experiments on the PASCAL VOC'07 dataset demonstrate that the proposed mHDSC outperforms mLDSC and other related sparse coding algorithms.

The proposed mHDSC is an implementation of the multiview learning, which has been widely applied in practical problems, such as cartoon synthesis [25,26] and cartoon correspondence construction [27]. In the future, we will apply the mHDSC for practical implementation including cartoon retrieval and cartoon classification.

Multiview Hessian Discriminative Sparse Coding for Image Annotation

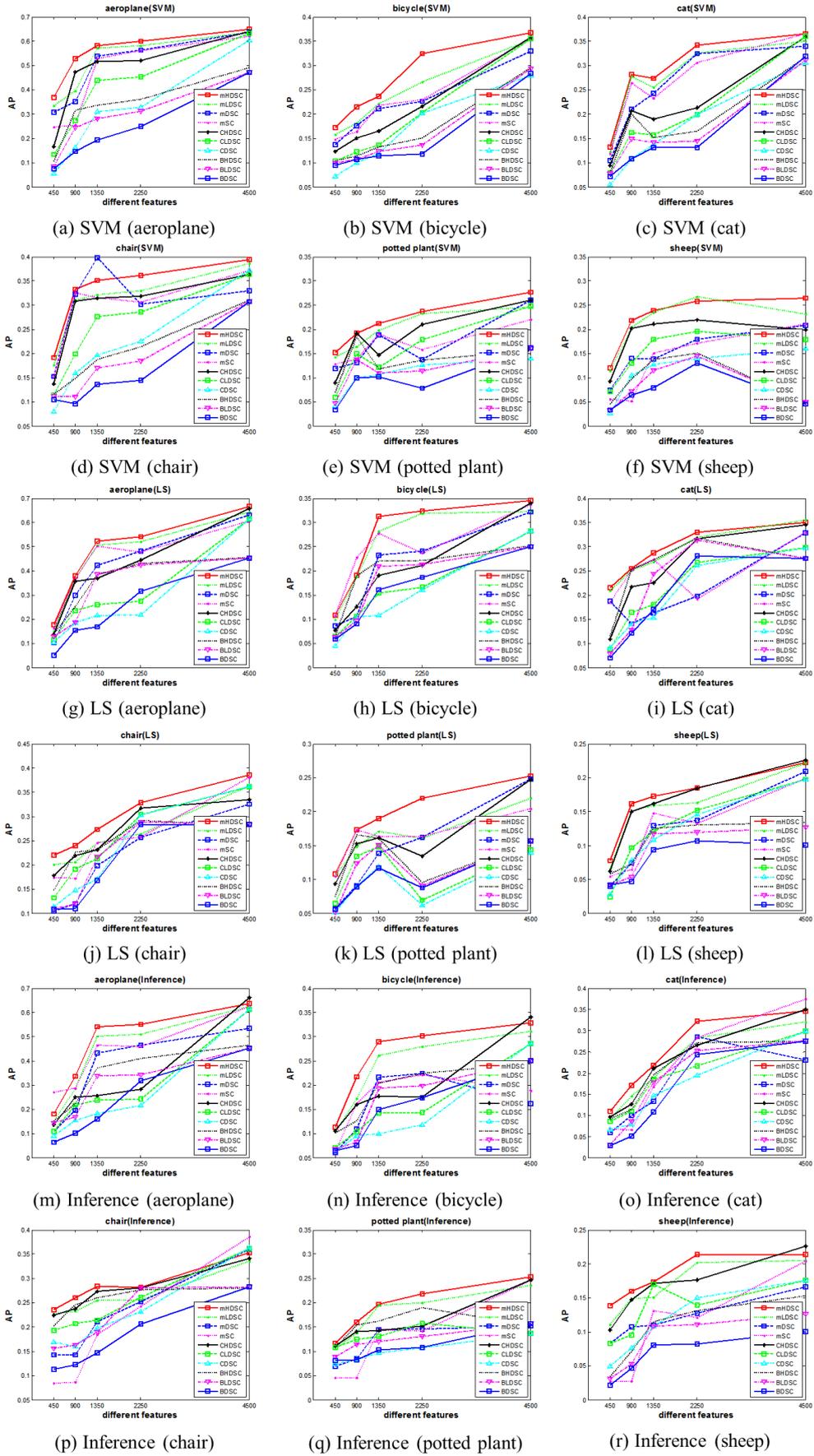

Fig. 6. The AP of different methods on selected visual classes.

**Figure Captions**

Fig. 1. mHDSC framework.

Fig. 2. Example images of the PASCAL VOC'07 database.

Fig. 3. The mAP of different views.

Fig. 4. The AP of different views on selected classes.

Fig. 5. The mAP of different methods.

Fig. 6. The AP of different methods on selected visual classes.

Multiview Hessian Discriminative Sparse Coding for Image Annotation

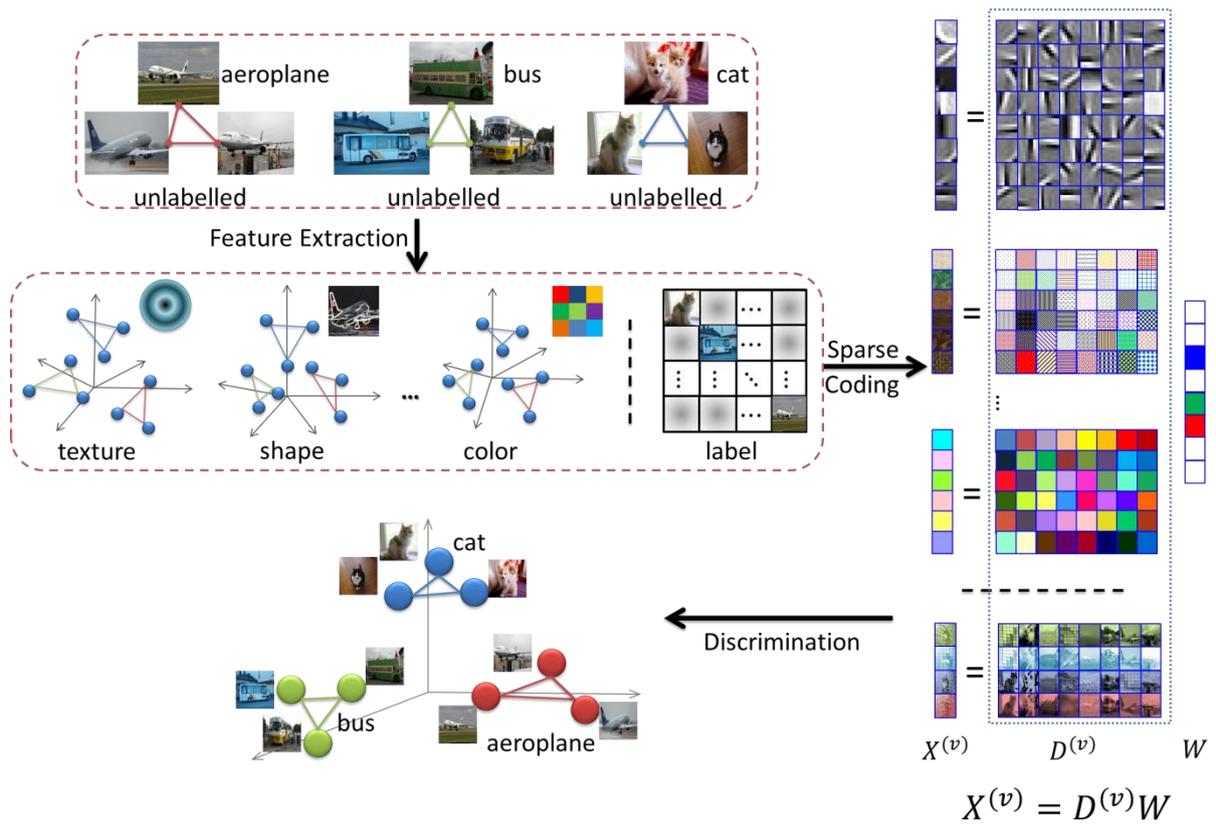

Fig. 1. mHDSC framework.

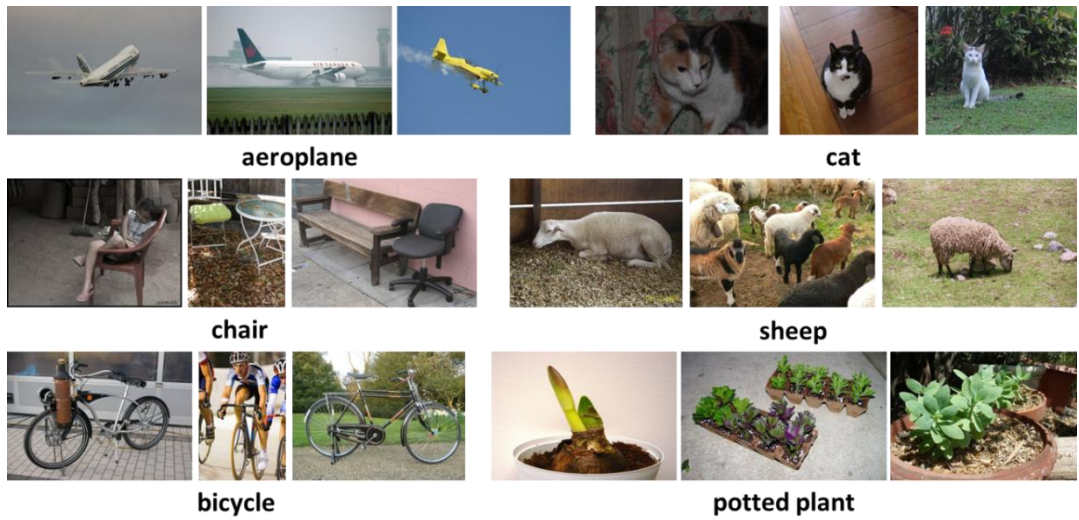

Fig. 2. Example images of the PASCAL VOC'07 database.

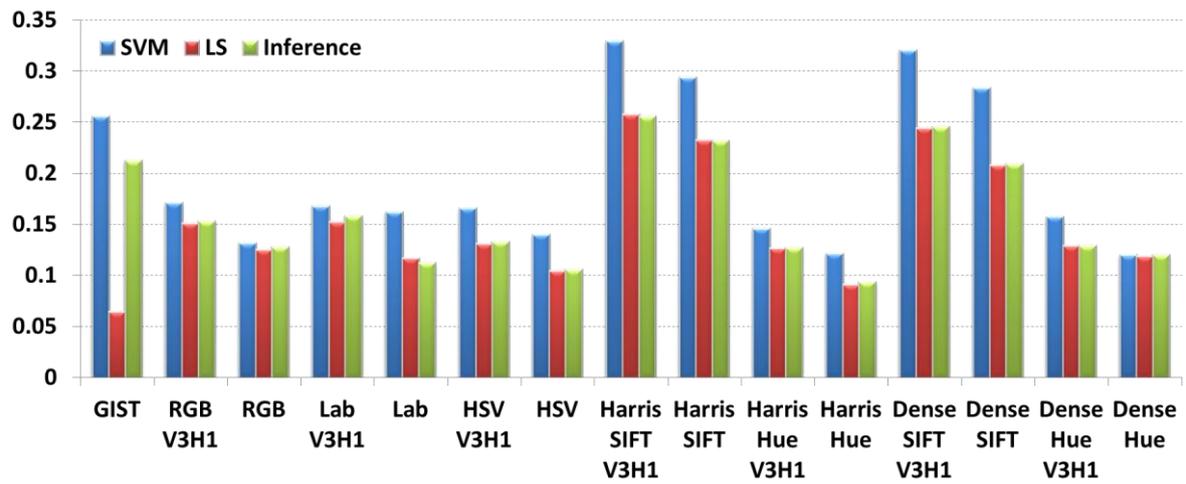

Fig. 3. The mAP of different views.

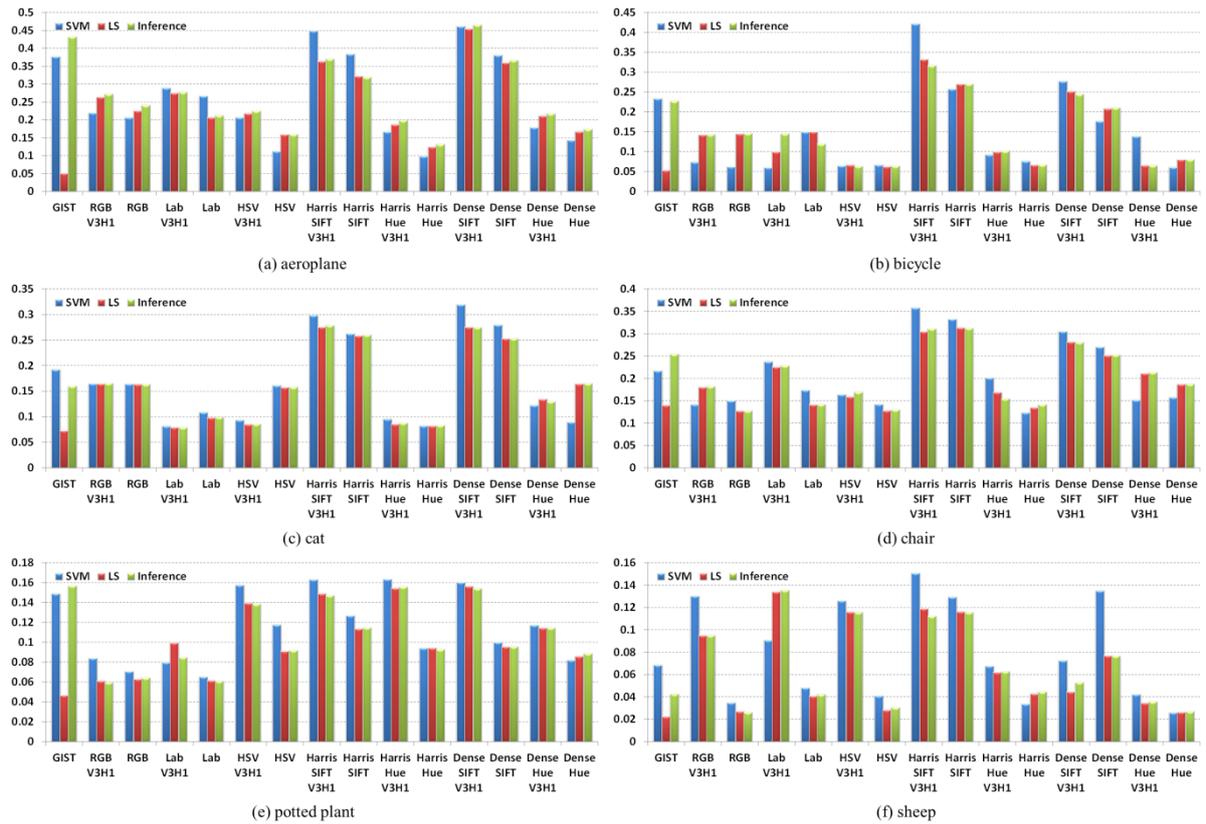

Fig. 4. The AP of different views on selected classes.

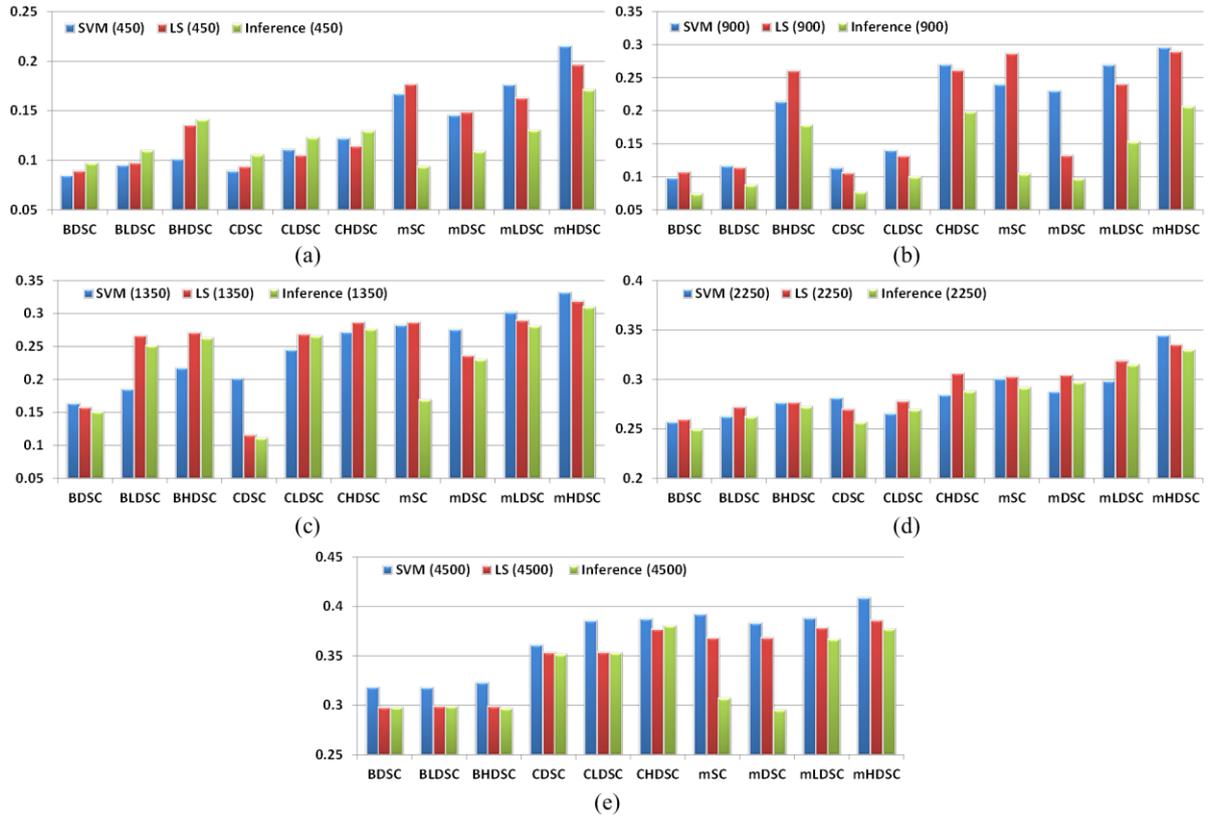

Fig. 5. The mAP of different methods.

Multiview Hessian Discriminative Sparse Coding for Image Annotation

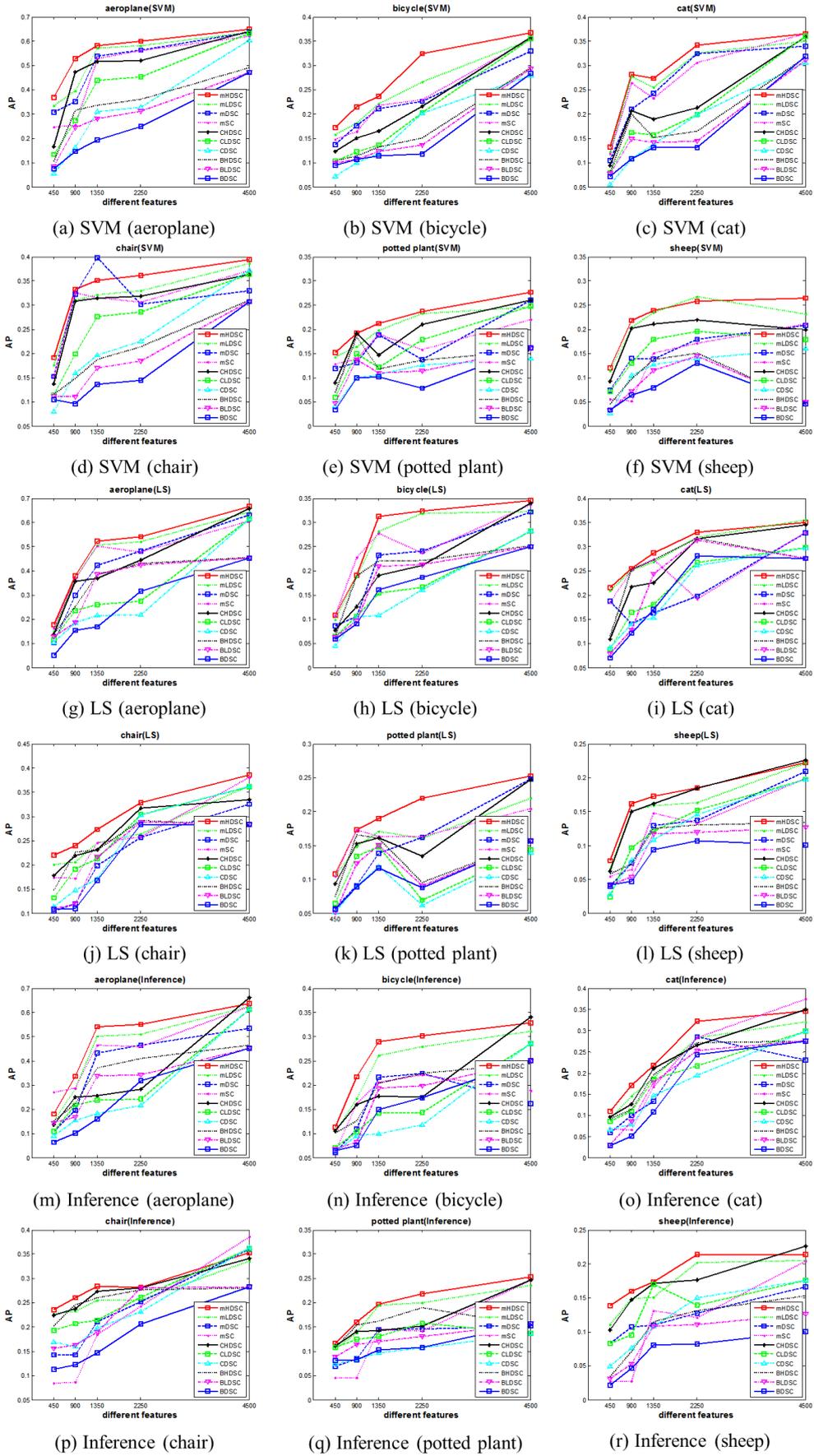

Fig. 6. The AP of different methods on selected visual classes.